\documentclass[11pt]{article}
\RequirePackage{amsthm,amsmath,amsfonts,amssymb, bm}
\RequirePackage{natbib}
\usepackage{xr-hyper}
\usepackage{longtable}
\RequirePackage[colorlinks,citecolor=blue,urlcolor=blue]{hyperref}
\RequirePackage{graphicx}
\RequirePackage[OT1]{fontenc}
\usepackage{color}
\usepackage{booktabs}
\usepackage{algorithm}
\usepackage[noend]{algpseudocode}
\usepackage{subcaption}
\usepackage{float}
\usepackage{caption}
\usepackage{algorithm,algcompatible,amsmath}
\usepackage{ragged2e}
\usepackage{multirow}
\usepackage{bbm}
\algnewcommand\INPUT{\item[\textbf{Input:}]}%
\algnewcommand\OUTPUT{\item[\textbf{Output:}]}%

\newcommand{\abs}[1]{\left\vert#1\right\vert}

\numberwithin{equation}{section}
\theoremstyle{plain}

\usepackage{authblk}

\begin{document}

\title{Are Made and Missed Different? An analysis of Field Goal Attempts of Professional Basketball Players via Depth Based Testing Procedure}
\author[a]{Kai Qi}
\author[b]{Guanyu Hu \thanks{Corresponding author: guanyu.hu@uth.tmc.edu}}
\author[c]{Wei Wu}
\affil[a]{Microsoft}
\affil[b]{Department of Biostatistics and Data Science, The University of Texas Health Science Center at Houston}
\affil[c]{Department of Statistics, Florida State University}
%\thankstext{T1}{Footnote to the title with the ``thankstext'' command.}

\date{}
\maketitle

\begin{abstract}
In this paper, we develop a novel depth-based testing procedure on spatial point processes to examine the difference in 
made and missed field goal attempts for NBA players.  Specifically, our testing procedure can statistically detect the differences between made and missed field goal attempts for NBA players. We first obtain the depths of two processes under the polar coordinate system. A two-dimensional Kolmogorov–Smirnov test is then performed to test the difference between the depths of the two processes. Throughout extensive simulation studies, we show our testing procedure with good frequentist properties under both null hypothesis and alternative hypothesis. A comparison against the competing methods shows that our proposed procedure has better testing reliability and testing power. Application to the shot chart data of 191 NBA players in the 2017-2018 regular season offers interesting insights about these players' made and missed shot patterns.  
\end{abstract}

\section{Introduction}\label{sec:intro}
Do made and missed field goal attempts follow different spatial processes for Stephen Curry? Does LeBron James have more field goal attempts in the region where he has a higher field goal percentage? In NBA, analyzing field goal attempts for different players is a crucial problem for each team. From the players' point of view, it helps them optimize their field goal attempts and design their training plans to improve their shooting skills. The optimal selection of field goal attempts for a player is that he or she can make field goal attempts on all locations without differences. From the teams' perspective, it helps them improve their training plans for different players or different positions. In order to answer the questions above, we need to know if there is any difference between the made and missed locations of field goal attempts. This paper aims to provide a statistical testing procedure to discriminate made and missed processes.

Quantitative analytics has been a key driving force in advancing modern professional sports, and there is no exception for professional basketball \citep{kubatko2007starting}. In NBA, shot chart data that contain both the location and the result of each field goal attempt will offer important insights about players' attacking styles and shed light on the evolution of defensive tactics. 
There have been many literature discussing shot chart data \citep[see,e.g.,][]{brian06, miller2014factorized,shortridge2014creating,gomez2015analysis,ervculj2015basketball,jiao2019bayesian, hu2020bayesiangroup, hu2020zero}. Most existing studies rely on developing statistical models to analyze the spatial patterns of different players' field goal attempts or find the latent subgroups among different players. \citet{brian06} proposed a hierarchical regression model under point-reference data framework for both field goal attempt frequency and accuracy over grids on the basketball court. 
However, they failed to capture the randomness of field goal attempts since the shot locations were fixed in their study. Considering randomness of field goal attempts in basketball game is very important for quantitative analysis of shot charts data. Compared to traditional point reference analysis in \citet{brian06}, spatial point processes \citep{diggle2013statistical} are natural to model randomness of locations. There is a large body of literature in the field of spatial point pattern data \citep[see,e.g.,][]{illian2008statistical, diggle2013statistical, guan2006composite, guan2010weighted, baddeley2017local, geng2019bayesian, jiao2020heterogeneity} under both parametric and nonparametric frameworks, which provides throughout the analysis of first-order (intensity) and second-order (pair correlation) properties for spatial point processes. \citet{miller2014factorized,yin2020bayesian} analyzed the intensity surface of field goal attempts of professional players, which offered interesting insights about the shooting patterns. Based on intensity estimates, \citet{hu2020bayesiangroup,yin2020analysis} discovered latent subgroups among NBA players to gain important insight about shooting similarities between different players. \citet{jiao2019bayesian} proposed a Bayesian marked point process framework for jointly modeling shot frequency and accuracy to discover the relationship between them. Although quantitative analysis (regression analysis and clustering analysis) of shot charts based on spatial point process has been well explored, discussions of the testing procedure for two or more processes are minimal.

The initial problem for testing is to select a proper statistical measure for the spatial point process. The statistical depth provides an effective framework to describe the central tendency \citep{liu2017generalized} and the ordered property \citep{qi2019dirichlet} for the spatial point process.  The notion of statistical depth was first introduced \citep{tukey1975mathematics} as a tool for visualizing bivariate data sets and has been extended to multivariate data over the last few decades. The depth is a measure of the centrality of a point with respect to a certain data cloud, which helps to set up center-outward ordering rules of ranks. Based on different centrality criteria, a large class of depths has been proposed, including the halfspace depth \citep{tukey1975mathematics}, convex hull peeling depth \citep{barnett1976ordering}, simplicial depth \citep{liu1990notion}, $L_1$-depth \citep{vardi2000multivariate}, and projection depth \citep{zuo2003projection}. The concept of statistical depth has been widely applied in outlier detection \citep{donoho1992breakdown}, multivariate density estimation \citep{fraiman1997multivariate}, non-parametric description of multivariate distributions \citep{liu1999multivariate}, and depth-based classification and clustering \citep{christmann2002classification}.

% Statistical depth has been extended to functional observations over the past two decades.  Band depth and half-region depth \citep{lopez2009concept, lopez2011half} are two commonly used functional depth, which have been successfully used for practical tasks such as classification.  Based on the graph representation, a number of extensions, modifications and generalizations have emerged such as local band depth \citep{agostinelli2013ordering}, set band depth \citep{whitaker2013contour}, spatially weighted band depth \citep{balzanella2015depth}, and empirical halfspace depth \citep{einmahl2015bridging}. Moreover, 
% functional extensions of the so-called spatial depth were studied in \citep{chakraborty2014spatial, sguera2014spatial, serfling2017depth}. This spatial depth is able to tackle scenarios where the solution of the problem is not extremely graphically clear.  \cite{narisetty2016extremal} introduced a notion called extremal depth, which satisfies the desirable properties of convexity and ``null at the boundary'', for which integrated data depth and band depth lack. Based on an elastic-metric-based measure of centrality for functional data,  \cite{cleveland2018robust} adopted band depth and modified band depth to estimate the template for functional data with phase variability. 

On the other hand, the depth-based statistical tests are still under-explored in both theory and applications. The earliest related work in this area, though different in the objective, was the bivariant rank test based on Jia's concept of ordering \citep{oja1983, oja1989, oja1992}. Then for multivariate data, a Wilcoxon's type of test is proposed by \citet{liu1993} (henceforth, Liu-Singh rank sum test) since the data depth is naturally related to ranking. As a theoretical foundation, \citet{zuo2006} elaborately studied the limiting distribution and asymptotic of Liu-Singh's rank sum test.  Multivariate depth functions can transfer data into a one-dimensional space, and many one-dimensional testing techniques can then be applied to the depth values. In this respect, a well-composed generalization can be found in \citet{zhang2012}. Another graphic form of depth-based test, the DD (depth vs. depth) plot, was introduced by \citet{liu1999multivariate} as an example of the potential application of depth function. Then \citet{li2004} systematical studied the DD plot by examining its performance w.r.t. location shift and scale expansion or contraction. In this paper, we focus on depth-based testing procedures for the spatial point process to explore the difference between made and missed field goal attempts among different players in NBA. 

The contribution of this paper is three-fold. First, we introduce the two-dimensional depth calculation for the spatial point process under the polar coordinate. Compared with the Cartesian coordinate, the polar coordinate represents the field goal attempts more appropriately since the distance and angle are the two most important factors for shooting selections by professional players \citep{brian06}. Then, we propose a new depth-based Kolmogorov–Smirnov test for two spatial point processes based on a two-dimensional depth. 
% To the best of our knowledge, this is the first attempt to apply a depth-based testing procedure to the spatial point process. The empirical studies validate the superior performance of our proposed testing procedure. Finally, our proposed testing procedure reveals several interesting findings on the NBA shot chart data.

The rest of this article is organized as follows. In Section \ref{sec:data}, we select one representative player for each position from the 2017–2018 NBA regular season and present their shot charts. In Section~\ref{sec:model}, we first briefly discuss the spatial point process model and depth calculation for spatial point pattern data. Then, a depth-based testing procedure is proposed to discriminate the difference between made and missed processes. Extensive simulation studies are conducted in Section~\ref{sec:simu} to investigate the empirical performance of the proposed testing procedure. Applications of the proposed testing procedure to 191 NBA players in the 2017–2018 regular season are reported in Section~\ref{sec:app}. We conclude this paper with a brief discussion in Section~\ref{sec:disc}.

\section{Motivating Data}\label{sec:data}
\begin{figure}[h]
\centering
\includegraphics[width = \textwidth]{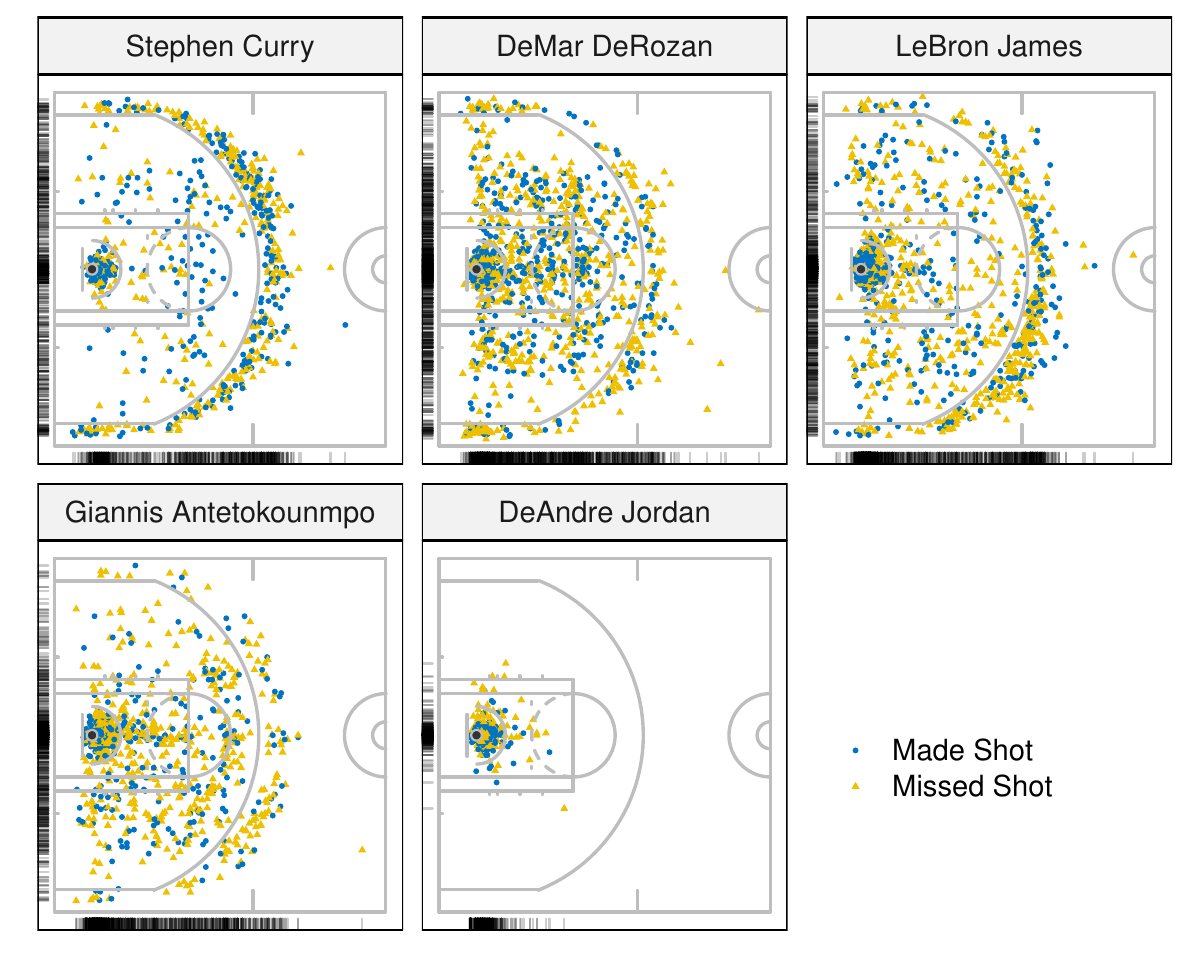}
\caption{Field Goal Attempts Data Display. On half court image, each point represents
one shot. The marginal distributions of the shot locations are shown on the left and down sides of each plot, respectively. 
(Players: Stephen Curry, DeMar DeRozan, LeBron James, Giannis Antetokounmpo, DeAndre Jordan)}
\label{fig:EDA}
\end{figure}

The NBA shot charts are freely available on the official NBA site \url{stats.nba.com}, and our study focuses on the regular season in 2017-2018. The data for each player contains information about all field goal attempts in the regular season. From the data, the half-court is positioned on a Cartesian coordinate system centered at the center of the rim, with $x$ ranging from $-250$ to 250 and $y$ ranging from $-50$ to 420, both with the unit of 0.1 foot, as the size of an actual NBA basketball half-court is $50 \ \text{feet} \times 47 \ \text{feet}$. 
We pick up one representative player for each position (Point Guard (PG), Shooting Guard (SG), Small Forward (SF), Power Forward (PF), or Center(C)), such as Stephen Curry, DeMar DeRozan, LeBron James, Giannis Antetokounmpo, and DeAndre Jordan. Figure~\ref{fig:EDA} shows their field goal attempts' locations, colored to indicate made or missed for each attempt.

As we can see from the plots, DeAndre Jordan does not have any field goal attempts from 3-point line. Most of the shots of Stephen Curry are made either close to the rim or right outside the arc (i.e., 3-point line). DeMar DeRozan tends to have a uniformly distributed shot pattern throughout the court. LeBron James has relatively more shots in the right-hand side area toward the basket. Giannis Antetokounmpo apparently has more missed attempts from 3-point line.
%\rc{comment out this sentence? The partial reason is that it is a more efficient strategy for players to pursue higher success rates near the rim or go after higher rewards by making 3-point shots.} 

\section{Model setup}\label{sec:model}
\subsection{Spatial Point Process}
For a particular player, the observed shot chart can be represented by two spatial point processes $\mathbf{S}_{made}=(\bm{s}_1,\bm{s}_2,\ldots,\bm{s}_{n_1})$ and $\mathbf{S}_{missed}=(\bm{s}'_1,\bm{s}'_2,\ldots,\bm{s}'_{n_2})$, where $\mathbf{S}_{made}$ and $\mathbf{S}_{missed}$ are the collections of the locations of made and missed field goal attempts, and $n_1$ and $n_2$ are the numbers of shots for made and missed processes, respectively.  

Spatial point process model provides a natural framework for capturing the random behavior of event location data. For example, in the made process, let $\bm{s}_i = (x_i, y_i), i=1,\ldots,n_1$ be the set of $x$- and $y$-coordinate locations for points that are observed in a pre-defined, bounded region $\mathcal{B} \subseteq \mathbb{R}^{2}$, which is the half basketball court. We denote the underlying stochastic mechanism that gives rise to the observed point pattern $\mathbf{S}_{made}$ as the spatial point process $\mathbf{Y}$. The process $N_{\mathbf{Y}}(A) = \sum_{i=1}^{n_1} \mathbbm{1}(\bm{s}_{i} \in A)$ is a counting process associated with the spatial point process $\mathbf{Y}$, which counts the number of points that fall into the area $A \subseteq \mathcal{B}$.  

A large body of literature on probability distributions exists for spatial point processes (see, e.g., \citet{diggle2013statistical}, and references therein), including Poisson processes, Gibbs processes, and Cox processes, etc.. In those literature, the most widely adopted class of models is the nonhomogeneous Poisson processes (NHPP), which assumes conditionally independent event locations with a deterministic intensity $\lambda(\mathbf{s})$. Under the NHPP framework, the number of events in area $A$, $N_{\mathbf{Y}}(A)$, follows a Poisson distribution with rate parameter $\Lambda(A) = \int_{A} \lambda(\mathbf{s}) d\mathbf{s}$. In addition, $N_{\mathbf{Y}}(A_{1})$ and $N_{\mathbf{Y}}(A_{2})$ are independent if two areas $A_1 \subseteq \mathcal{B}$ and $A_2 \subseteq \mathcal{B}$ are disjoint. Given the observed point pattern $\mathbf{S}_{made}$ on fixed region $\mathcal{B}$, the likelihood of NHPP is 
\begin{equation*}
\label{eq:NHPP_lik}
L = \frac{\prod_{i=1}^{n_1} \lambda(\mathbf{s}_{i})}{\exp(\int_{\mathcal{B}} \lambda(\mathbf{s}) d\mathbf{s})},
\end{equation*}
where $\lambda(\mathbf{s}_{i})$ is the intensity function evaluated at location $\mathbf{s}_{i}$. As mentioned in Section~\ref{sec:intro}, most existing literature focuses on intensity estimation or pair correlation estimation and joint model for marked point process. In this paper, our goal is to propose a testing procedure for the null hypothesis
\begin{equation*}
    H_0:\quad \mathbf{S}_{made}\text{ and }\mathbf{S}_{missed} \text{ are the same process}. 
\end{equation*}
Our testing procedure is based on the depth of spatial point process.

\subsection{Depth Calculation}
Let $\mathbf{S}$ denotes a collection of locations in a bounded set $\mathcal{B}\subseteq \mathbb{R}^2$, a depth function essentially is a map from $\mathcal{B}$ to $\mathbb{R^+}$ which generates an appropriate center-outward rank for locations of points.

Depth functions for multivariate data can be broadly categorized as nonparametric depth (e.g., half-space/Tukey depth) and parametric forms (e.g., Mahalanobis depth). While different depth functions have various properties (such as shapes and contours), one should choose an appropriate depth based on the nature of data. Here we briefly describe two classical depth functions as follows.

Let $\mathbf{P}$ be the probability measure on $\mathbb{R}^d$, $d\geq1$, then for a point $z\in \mathbb{R}^d$, the Mahalanobis depth and Tukey's depths are defined in the following forms:

\textbf{Mahalanobis Depth (MD)} \citep{liu1993}: let $\mu$ and $\Sigma$ denote the mean vector and variance-covariance matrix of $\mathbf{P}$, the Mahalanobis depth (MD) for $z$ w.r.t. $\mathbf{P}$ is
\begin{equation}
\label{MD}
MD(z;\mathbf{P})=[1+(z-\mu)^{T}\Sigma^{-1}(z-\mu)]^{-1}.
\end{equation}

\textbf{Half-space/Tukey's Depth (HD)} \citep{tukey1975mathematics}: The HD of $z$ w.r.t. $\mathbf{P}$ is defined as
\begin{equation}
\label{HD}
HD(z;\mathbf{P})=\inf_{H}\{P(H): \text{H is a closed halfspace in } \mathbb{R}^d \text{ containing z}\}.
\end{equation}

By replacing $\mathbf{P}$ in Equations \eqref{MD} and \eqref{HD} with empirical distribution $\mathbf{P}_n$, we have the sample version of Mahalanobis depth and Tukey's depth. Both Mahalanobis depth and Tukey's depth for multivariate data are well established. In contrast to the stereotyped elliptic depth contours of Mahalanobis depth, Tukey's depth provides more flexible and appropriate contours. As a nonparametric depth function, Tukey's depth does not rely on the estimation of mean and variance. In the following sections, we will adopt Tukey's depth except otherwise specified. 

For $d>1$, the computation relies on effectively finding a hyperplane containing $z_i$ that satisfies the definition of Tukey's depth in Equation \eqref{HD}.  Such computation can be extremely inefficient or infeasible, especially for high-dimensional data.
In this paper, the computation of depth has never been extended to more than two dimensions. For $d=2$, a straightforward algorithm for Tukey's depth needs $\mathcal{O}(n^2)$ steps, we will adopt a faster algorithm proposed by \cite{peter96} which only requires $\mathcal{O}(n\log{}n)$ steps, where $n$ denotes sample size. 
%Such an algorithm has been incorporated into the R package "depth", and in this paper, we will adopt the "depth" package to compute Tukey's depth for $d=2$.

\subsection{Depth Based Testing}
\label{sec:depth_tests}
The performance of parametric statistical tests often relies on the normality assumption of the underlying distribution.  In contrast, some nonparametric tests can perform better when dealing with symmetrically distributed data (e.g. KS test \citep{Mohd2011}, Wilcoxon's test). To address this, a data transformation function is needed to improve the reliability and consistency of tests.  On the other hand, it is generally challenging to extend the one-dimensional tests to multivariate space directly. For example, since the maximum difference of two joint cumulative distribution functions is not clearly defined, the Kolmogorov-Smirnov test (KS test) cannot be simply generalized for high dimensional data. One common approach to generalizing those tests to high dimensional space is through dimensional reduction techniques.

In general, a multivariate depth function is a continuous function mapping data from $\mathbb{R}^d$ to $\mathbb{R^+}$ that holds the following four properties \citep{zuo2000general}: affine invariance, maximality at the center, monotonicity relative to the deepest point, and vanishing at infinity. All these properties make multivariate depth function an ideal pre-defined transformation when performing statistical tests on high-dimensional data.

Here, we briefly review the depth-based goodness-of-fit tests that treat depth functions as transformations to reduce data dimensionality, include Liu-Singh rank-sum test, depth-based Pearson's chi-square test, and depth-based KS test and Cramer-Von Mises test. We will then propose our new depth-based test.

\subsubsection{Liu-Singh rank sum test}
%A depth function is designed to measure the center-outward order and generate center-outward rank for a given type of data. Therefore, 
The well-known Wilcoxon rank sum can naturally incorporate depth functions. In this respect, \citet{liu1993,zuo2006} systematically generalized the Wilcoxon rank sum test to multivariate data through depth functions (refer to as Liu-Singh depth-based rank sum test). Briefly, the Liu-Singh depth-based rank sum test can be described as:

Let $X \sim H$ and $Y \sim G$ be two independent random variables in $\mathbb{R}^d$, and $D(\cdot;H)$ be a depth function of a given distribution $H$ that maps $\mathbb{R}^d$ to $[0,1]$. Then the outlyingness of y w.r.t. $H$ can be measured by $R(y;H)=P_H(X: D(X;H)\leq D(y;H))$, and we define:
$$Q(H, G):=\int R(y;H) dG(y)=P\{D(X;H)\leq D(Y;H) \mid X \sim H, Y \sim G\}.$$
Following Proposition 3.1 in \citet{liu1993}, for continuously distributed $D(\cdot;H)$, $R(Y;H)$ follows a uniform distribution in $[0, 1]$. Under the null hypothesis $H_0: H=G$,  the test statistic $Q(H, G)= \frac{1}{2}$ and $Q(H,G)<\frac{1}{2}$ indicates the possible location shift or scale increase from $H$ to $G$. 

In practice, one can compare two empirical distributions $H_m \text{ and } G_n$ through the two-sample version of Liu-Singh depth-based rank sum test, such that:
\begin{equation}
\label{Liu-singh test}
 Q(H_m,G_n):=\int R(y; H_m) dG_n(y)=\frac{1}{n}\sum_{j=1}^{n}R(Y_j;H_m)   
\end{equation}

While the asymptotic results rely on the dimensionality of data and properties of depth function. For dimension $d=1$, under null hypothesis, \citet{liu1993} has shown that:
$$((1/m +1/n)/12)^{-1/2}(Q(H_m,G_n)-1/2)\rightarrow N(0,1).$$
By assuming more conditions on the depth function, \citet{zuo2006} extended the asymptotic studies of Liu-Singh depth-based rank sum test to higher dimensions under both null and alternative hypothesis.

\subsubsection{Depth-based Pearson's chi-square test}
Another important depth-based test is derived from Pearson's chi-square test, which involves discretizing and separating the continuous depth values into multiple intervals. By counting the number of observations that fall into those predefined depth value intervals, one can form a classical Pearson's chi-squared test against expected frequencies. Here we briefly describe the test procedure as follow:

Let $X\sim F$ be a random variable in $\mathbb{R}^d$ and $D(\cdot;F)$ be a depth function that maps $X$ to [0, 1]. Given a random sample of size $n$, under $H_0: F=F_0$, denote $z_{j,F_0}=D(x_j; F_0)$ as the depth of $x_j$ with respect to $F_0, j=1,2,\cdots,n$. Assuming the depth values can be separated by $a_i, i=1,\cdots,r$, such that:
$$0=a_0<a_1<\cdots<a_{r-1}<a_r=1.$$
Define
$$A_i=\{ X : a_{i-1} < D(X ; F_0)\leq a_i, X \in R^d\}, i=1,\cdots, r.$$
$$p_i=P_{F_0}(A_i), \ n_i=\# \{ j: a_{i-1} < z_{j,F_0} \leq a_i, j= 1,\cdots, n \}, i=1,\cdots,r.$$

Now, we have the expected frequency $np_i$ and observed frequency $n_i$ for $r$ regions: $A_1, A_2, \cdots, A_r \in \mathbb{R}^d$, and the Pearson's chi-square test statistic shown in Section 2.1 in \cite{zhang2012} is computed by
\begin{equation}
\label{chi-square test}
\chi^2=\sum_{i=1}^r \frac{(n_i-np_i)^2}{np_i} \sim \chi^2(r-1). 
\end{equation}

To make the depth-based chi-square test powerful, one does not only require the sample size $n \geq 50$, but also need to ensure the theoretical frequencies $np_i \geq 5$ for each segmented region $A_i, i=1,\cdots, r$. 
%%%%%
\subsubsection{Depth-based Cramer-Von Mises and Kolmogorov-Smirnov test}
\label{sec:cm and ks}
A multivariate depth function transforms data from high-dimensional space to one dimension. Therefore, some general one-dimensional tests, such as the classical Kolmogorov-Smirnov (KS) test and Cramer-Von Mises (CM) test, can be performed on the transformed data.   

Denote $F_0(x)$ as a continuous distribution function and $F_n(x)$ as the empirical distribution function of sample $X_1, X_2, \cdots, X_n \in \mathbb{R}^d$. Then the KS and CM statistics can be denoted as:
$$KS=\sup_{x\in R^d}|F_n(x)-F_0(x)|, \ CM=\int_{R^d} [F_n(x)-F_0(x)]^2 dF_0(x)$$

The KS test or CM test depends largely on the distribution $F_0$, and technically, they are not nonparametric tests. But one can apply a depth-based transformation $R(\cdot)$ \citep{zhang2012} of Liu-Singh’s test as
$$T_i=P_{F_0}(X:D(X ; F_0)\geq D(X_i; F_0)), i=1,2,\cdots,n.$$

Under Proposal 3.1 in \cite{liu1993}, $T_i$ follows a uniform distribution in $[0,1]$ under null hypothesis. The test of equal distribution is equivalent to: $H_0: T_1, \cdots T_n$ is a  random sample from $U[0,1]$. Denote $G_n(t)$ as the empirical distribution function of  $\{T_i, i=1,\cdots,n\}$, then the one-dimensional KS test statistic and CM test statistic based on $G_n(t)$ are:
$$KS^{*}=\sup_{t\in [0,1]}|G_n(t)-t|, \ CM^{*}=\int_0^1[G_n(t)-t]^2dt.$$
Notice that both test statistics $KS^{*}$ and $CM^{*}$ do not depend on $F_0$, and they are nonparametric tests.

\subsection{Two-dimensional KS test}
The depth-based statistical tests mentioned above include two essential steps: 1) apply a multivariate depth function to reduce the dimensionality of each data point to one, and 2) perform a one-dimensional test on the scalar depth values. Inevidiblely, part of dimensional information (e.g., shape) is lost during the dimension reduction. In this section, we seek an alternative approach of incorporating depth into hypothesis testing.

Among many popular multivariate tests of equal distribution, it is well known that the KS test is sensitive to location shift and shape change. 
%Although various studies have indicated that the KS test is less powerful for testing normality than the Shapiro-Wilk test or Anderson-Darling test, it performs better when testing for uniformity \citep{stephens74}. As we have mentioned in Section \ref{sec:cm and ks}, if we define a depth-based transformation as
%$$T_i=P_{F_0}(X:D(X ; F_0)\geq D(X_i; F_0)), i=1,2,\cdots,n,$$
%for $X \in R$ and $D(\cdot)$ being continuous, then the statistic $T_i$ will follows a uniform distribution in $[0,1]$. 
% On the other hand,
In practice, the generalization of the KS statistic to multivariate settings is challenging, since the maximum difference between two joint cumulative distribution functions is not well defined. One solution proposed by \cite{ana97} is Rosenblatt's transformation, but the computational process of Rosenblatt's transformation is rather complicated to use.  Another natural approach is to compare the CDFs of two samples from all possible orders. In this aspect, \cite{Peacock83} proposed a two-dimensional KS test which is hard to extend to more than two dimensions due to the computational cost. \cite{fasano87} further improved Peacock's work by proposing a faster version which also considers the correlation of data and sample size, and then generalized it to a three-dimensional case. Although the theoretical study of Peacock's test for dimensionality beyond three is lacking, we mainly focus on two-dimensional data in this paper.

For dimension $d=2$, Peacock’s method compares the integrated probability in each of four natural quadrants around a given point $(x_i, y_i)$, namely the total probabilities in $(x > x_i , y > y_i), (x < x_i, y > y_i), (x < x_i, y < y_i), (x > x_i, y < y_i)$. Then the two-dimensional KS statistic $T_{KS}$ is the maximum difference of the corresponding integrated probabilities over all data points and all quadrants.

Note that the distribution of statistic $T_{KS}$ in general may not be independent of the shape of underlying two-dimensional distribution under the null hypothesis. %, the computation of P-value based on $T_{KS}$ seems infeasible. 
\citet{fasano87} approximated the distribution of $T_{KS}$ as a function of sample size and correlation through extensive Monte Carlo simulations. Furthermore, by studying the Monte Carlo simulation results, the significant level of the two-dimensional KS test can be approximated by
\begin{equation} \label{appoxD} 
Pr(T_{KS}>observed)=Q_{KS}(\frac{\sqrt{N} T_{KS}}{1+ \sqrt{1-r^2}(0.25-0.75/\sqrt{N})}),
\end{equation}
where $N$ is the sample size, 
$$Q_{KS}(x)=2\sum_{j=1}^{\infty}(-1)^{j-1}e^{-2j^2x^2},$$ 
and $r$ is the correlation coefficient. For comparing two samples with sizes $n_1$ and $n_2$, approximation Equation \eqref{appoxD} still holds with $N=\frac{n_1 n_2}{n_1+n_2}$. Notice that, the above approximation formula is accurate enough when $N > 20$, and when the indicated probability (significance level) is less than 0.20.

\subsection{A new two-dimensional depth-based KS test}
\label{sec:proposed test}
We now turn to propose our new statistical test based on statistical depths for basketball players' shot charts. The statistical scheme we presented in this section is built upon the polar representation of shot locations. Polar coordinates have been widely applied in shooting data analysis \citep{brian06,Gudmundsson17}. During a basketball game (among many other games), players move around and focus on a fixed point (the basket); therefore, it is more natural to interpret a shot location as polar coordinates than the conventional Cartesian coordinate. 

Then, we are interested in testing if two shooting patterns are the same. Suppose one shooting pattern follows a Poisson process with intensity function $\lambda_1(\cdot)$ and another follows a spatial point process with intensity function $\lambda_2(\cdot)$. The goal of the proposed procedure is to test
$$H_0: \lambda_1(\cdot) = \lambda_2(\cdot).$$
Notice that their intensity function completely defines the two spatial point processes; therefore, the null hypothesis is equivalent to
$$H_0: F_1(\cdot) = F_2(\cdot),$$
where $F_1$ and $F_2$ are the corresponding distribution functions of two spatial point processes.

Suppose two independent spatial point process samples $\mathbf{S}_1=(s_1,\cdots,s_{n_1})$ and $\mathbf{S}_2=(s'_1,\cdots,s'_{n_2})$ have underlying intensity function $\lambda_1$ and $\lambda_2$, respectively, where $s_i=(x_i,y_i), \ i=1,\cdots,n_1$ and $s'_j=(x'_j,y'_j) , \ j=1,\cdots,n_2$ are the observed locations in the pre-defined, bounded region $\mathcal{B}\in \mathbb{R}^2$. Then the proposed test procedures can be list as following steps.
\begin{enumerate}
    \item Transfer Cartesian  coordinate $(x,y)$ to polar coordinate $(r, \theta)$ by:
$$r=\sqrt{(x^2+y^2)}, \theta = \arctan(\frac{y}{x}).$$
Denote the transformed $\mathbf{S}_1, \mathbf{S}_2$ as $\mathbf{S}^*_1=({s}^*_1,\cdots,{s}^*_{n_1})$ and $\mathbf{S}^*_2=({s}^{*'}_1,\cdots,{s}^{*'}_{n_2})$, where ${s}^*_i=(r_i, \theta_i)$ and ${s}^{*'}_j=(r'_j, \theta'_j)$. For basketball shot locations in this paper, we set the basket as the origin.
\item Estimated the pooled depth values for every location in $\mathbf{S}^*_1$ and $\mathbf{S}^*_2$. 

Under $H_0: \lambda_1=\lambda_2$, we merge two sets as
$$\mathbf{S}^*_{pool}=({s}^*_1,\cdots,{s}^*_{n_1},{s}^{*'}_1,\cdots,{s}^{*'}_{n_2}).$$

As discussed previously, instead of using a two-dimensional depth, we apply an one-dimensional halfspace depth function $D(\cdot)$ on $r$ and $\theta$ separately. For one-dimensional halfspace depth, $D(\cdot)$ is uniformly distributed in $[0,\frac{1}{2}]$ if the underlying distribution is continuous. Since the one-dimensional halfspace depth can be written as: $D(x)=\min(F(x),1-F(x))$, where $F(x)$ is the CDF a of continuous variable X. It is well known that $F(x)$ is uniformly distributed between $[0,1]$, hence $0 \leq D(x) \leq 1/2$, and:
\begin{align*}
P\{D(x) \leq d\}&=P\{F(x) \leq d, F(x) \leq 1/2\}+P\{1-F(x) \leq d, F(x) \geq 1/2 \} \\
&=2\cdot P\{ F(x) \leq d, F(x) \leq 1/2) \} = 2d.
\end{align*}

The pooled depth of location $(r,\theta)$ based on the pooled set therefore is a two-dimensional value $(D(r; \mathbf{S}^*_{pool}), D(\theta; \mathbf{S}^*_{pool}))$, where $D(r; \mathbf{S}^*_{pool})$ and $D(\theta; \mathbf{S}^*_{pool})$ measure the center-outward ranks of $(r,\theta)$ based on $\mathbf{S^*}_{pool}$ w.r.t. distance $r$ and angular $\theta$, respectively.  $D(r; \mathbf{S}^*_{pool})$ and $D(\theta; \mathbf{S}^*_{pool})$ can be calculated as

$$ \scriptstyle{D(r; \mathbf{S}^*_{pool})=\min(\frac{\#\{r_k \in \mathbf{S}^*_{pool}: r_k \leq r , k=1,\cdots,n_1+n_2\}}{n_1+n_2}, \frac{\#\{r_k \in \mathbf{S}^*_{pool}: r_k \geq r , k=1,\cdots,n_1+n_2\}}{n_1+n_2})},$$
and 
$$ \scriptstyle{D(\theta; \mathbf{S}^*_{pool})=\min(\frac{\#\{\theta_k \in \mathbf{S}^*_{pool}: \theta_k \leq \theta , k=1,\cdots,n_1+n_2\}}{n_1+n_2}, \frac{\#\{\theta_k \in \mathbf{S}^*_{pool}: \theta_k \geq \theta , k=1,\cdots,n_1+n_2\}}{n_1+n_2})}.$$

By applying above functions to $\mathbf{S}^*_1$ and $\mathbf{S}^*_2$, we get the two-dimensional depth representation for points in $\mathbf{S}^*_1$ and $\mathbf{S}^*_2$, denote as $\mathcal{D}_1$ and $\mathcal{D}_2$.
\item  Calculate the two-dimensional KS statistic $T_{KS}$ and approximate the P-value.

The two-dimensional KS statistic $T_{KS}$ is the maximum difference of the corresponding integrated probabilities over all data points and four natural quadrants. But when comparing two samples, $T_{KS}$ depends on which dataset ranged over. As proposed by \cite{press88}, one solution is to use the average of $T_{KS}$ values obtained from ranging two samples. 

For notation purpose, we denote $D(r; \mathbf{S}^*_{pool})$ as $d_r$, $D(\theta; \mathbf{S^*}_{pool})$ as $d_{\theta}$.   
Let $C^1_{\mathcal{D}_1}(d_r,d_\theta)$, $C^2_{\mathcal{D}_1}(d_r,d_\theta)$, $C^3_{\mathcal{D}_1}(d_r,d_\theta)$, $C^4_{\mathcal{D}_1}(d_r,d_\theta)$ denote the counts of points of set $\mathcal{D}_1=\{(d_{r_i},d_{\theta_i}),i=1,\cdots,n_1\}$ in quadrants $(d_r > d_{r_i} , d_\theta > d_{\theta_i})$, $(d_r \leq d_{r_i}, d_\theta > d_{\theta_i})$, $(d_r \leq d_{r_i}, d_\theta \leq d_{\theta_i})$, $(d_r > d_{r_i}, d_\theta \leq d_{\theta_i})$, respectively. Then, 
$$ T_{KS1}=\max_{(d_{r_i},d_{\theta_i}) \in \mathcal{D}_1} \left [
\abs{\frac{C^1_{\mathcal{D}_1}(d_{r_i},d_{\theta_i})}{n_1}-\frac{C^1_{\mathcal{D}_2}(d_{r_i},d_{\theta_i})}{n_2}}, 
\abs{\frac{C^2_{\mathcal{D}_1}(d_{r_i},d_{\theta_i})}{n_1}-\frac{C^2_{\mathcal{D}_2}(d_{r_i},d_{\theta_i})}{n_2}}, \right .$$
$$ \left. \abs{\frac{C^3_{\mathcal{D}_1}(d_{r_i},d_{\theta_i})}{n_1} - \frac{C^3_{\mathcal{D}_2}(d_{r_i},d_{\theta_i})}{n_2}},
\abs{\frac{C^4_{\mathcal{D}_1}(d_{r_i},d_{\theta_i})}{n_1}-\frac{C^4_{\mathcal{D}_2}(d_{r_i},d_{\theta_i})}{n_2}} \right ] $$

$$T_{KS2}=\max_{(d_{r'_i},d_{\theta'_i}) \in \mathcal{D}_2} \left [
\abs{\frac{C^1_{\mathcal{D}_1}(d_{r'_i},d_{\theta'_i})}{n_1}-\frac{C^1_{\mathcal{D}_2}(d_{r'_i},d_{\theta'_i})}{n_2}}, 
\abs{\frac{C^2_{\mathcal{D}_1}(d_{r'_i},d_{\theta'_i})}{n_1}-\frac{C^2_{\mathcal{D}_2}(d_{r'_i},d_{\theta'_i})}{n_2}}, \right . $$ 
$$ \left . \abs{\frac{C^3_{\mathcal{D}_1}(d_{r'_i},d_{\theta'_i})}{n_1} - \frac{C^3_{\mathcal{D}_2}(d_{r'_i},d_{\theta'_i})}{n_2}},
\abs{\frac{C^4_{\mathcal{D}_1}(d_{r'_i},d_{\theta'_i})}{n_1}-\frac{C^4_{\mathcal{D}_2}(d_{r'_i},d_{\theta'_i})}{n_2}} \right ]$$
and,
$$T_{KS}=\frac{T_{KS1}+T_{KS2}}{2}.$$
The p-value can be estimated using Equation \eqref{appoxD}. The detailed algorithm can be found in Section 14 of \cite{press88}. Once the p-value is calculated, one can compare it with the critical value (e.g., 5\%) to decide whether to reject the null hypothesis.
\end{enumerate}

Both proposed depth-based KS test and KS test we described in Section \ref{sec:cm and ks} assume that $\mathbf{S}_1$ and $\mathbf{S}_2$ are independent. Comparing with depth-based tests we discussed in Section \ref{sec:depth_tests}, the proposed framework is computationally efficient.  This is because one only needs to loop through four quadrants rather than all possible directions. The computation time can be further reduced by the approximation methods we described. Besides, we expect the proposed test to achieve a better performance since it utilizes more information to form test statistics. In the following section we will compare our proposed test with other tests on simulations and real data.

\section{Simulation Study}\label{sec:simu}
In this section, we mainly focus on testing the performance of our proposed test method through simulations. For comparing two spatial processes, an effective statistical testing scheme is the one that is sensitive to minor variations in distributions/intensities (i.e., power of the test), yet  insensitive to noises when the underlying distributions are the same (i.e., type I error). Therefore, the simulations in this section are designed to evaluate the performance of our proposed test w.r.t. controlling type I error and enhancing the power of the test. 

Our proposed test scheme consists of three components: adoption of the polar system, transformation through one-dimensional depth function, and a two-dimensional KS test. We want to test the role and performance of each component. Also, we would like to compare our proposed test scheme with other classical depth-based tests or traditional "goodness-of-fit" tests. For comparison, we will perform Liu-Singh’s rank sum test, the two-dimensional KS test, and our proposed test on the same simulated datasets, on both the original Cartesian system and polar system.

The detailed test schemes are listed as follows:
 \begin{itemize}
\item \textbf{Method 1}: Liu-Singh's rank sum test, as introduced in Section \ref{sec:depth_tests}. To be consistent, we adopt a two-dimensional Tukey's depth (Equation \ref{HD}) as the transformation depth function. 
Then by Corollary 2 and Theorem 1 in \cite{zuo2006}, assume $H$ is continuous, the following asymptotic result holds for Tukey's depth
$$(\sigma^2_{GH}/m +\sigma^2_{HG}/n)^{-1/2}(Q(H_m,G_n)-Q(H,G))\xrightarrow{d} N(0,1), as \ \min(m,n)\rightarrow \infty,$$
where under the null hypothesis $H=G$, $Q(H,G)=1/2$ and $\sigma^2_{GH} =\sigma^2_{HG}=1/12$. This implies that the p-value can be estimated by $N(0,1)$ according to the asymptotic result above.
%For comparing two empirical distribution $H_m$ and $G_n$, since the quality index $Q$ in Equation \ref{Liu-singh test} is not symmetric, we construct the test statistics as
%$$Q =\min(Q(H_m,G_n), Q(G_n,H_m)).$$ {\bf use another notation}
\item \textbf{Method 2}: A variant of Method 1 by adopting the polar coordinate. We introduce an additional transformation layer that maps data from Cartesian coordinates to polar coordinates before employing the two-dimensional Tukey's depth.  The purpose of this method is to isolate the impact of the polar system and evaluate it separately.

\item \textbf{Method 3}: The popular two-dimensional KS test. Since we have elaborately described the two-dimensional KS test in Section \ref{sec:proposed test}, we will omit the test details here.

\item \textbf{Method 4}: The two-dimensional KS test on the polar coordinate system. Similar to Method 2, we perform an two-dimensional KS test on the polar coordinate.

\item \textbf{Method 5}: Same as our proposed test scheme in Section \ref{sec:proposed test}, except that we omit the first step and directly work on the original Cartesian coordinates. That is, this method contains 2 steps: first estimate the pooled depth of location $(x, y)$, and then perform a two-dimensional KS test on pooled depth.

\item \textbf{Method 6}: Our proposed method as described in Section \ref{sec:proposed test}.
 \end{itemize}

\subsection{Performance on controlling type I error}

\begin{figure}[htp]
    \centering
    \includegraphics[width=\textwidth]{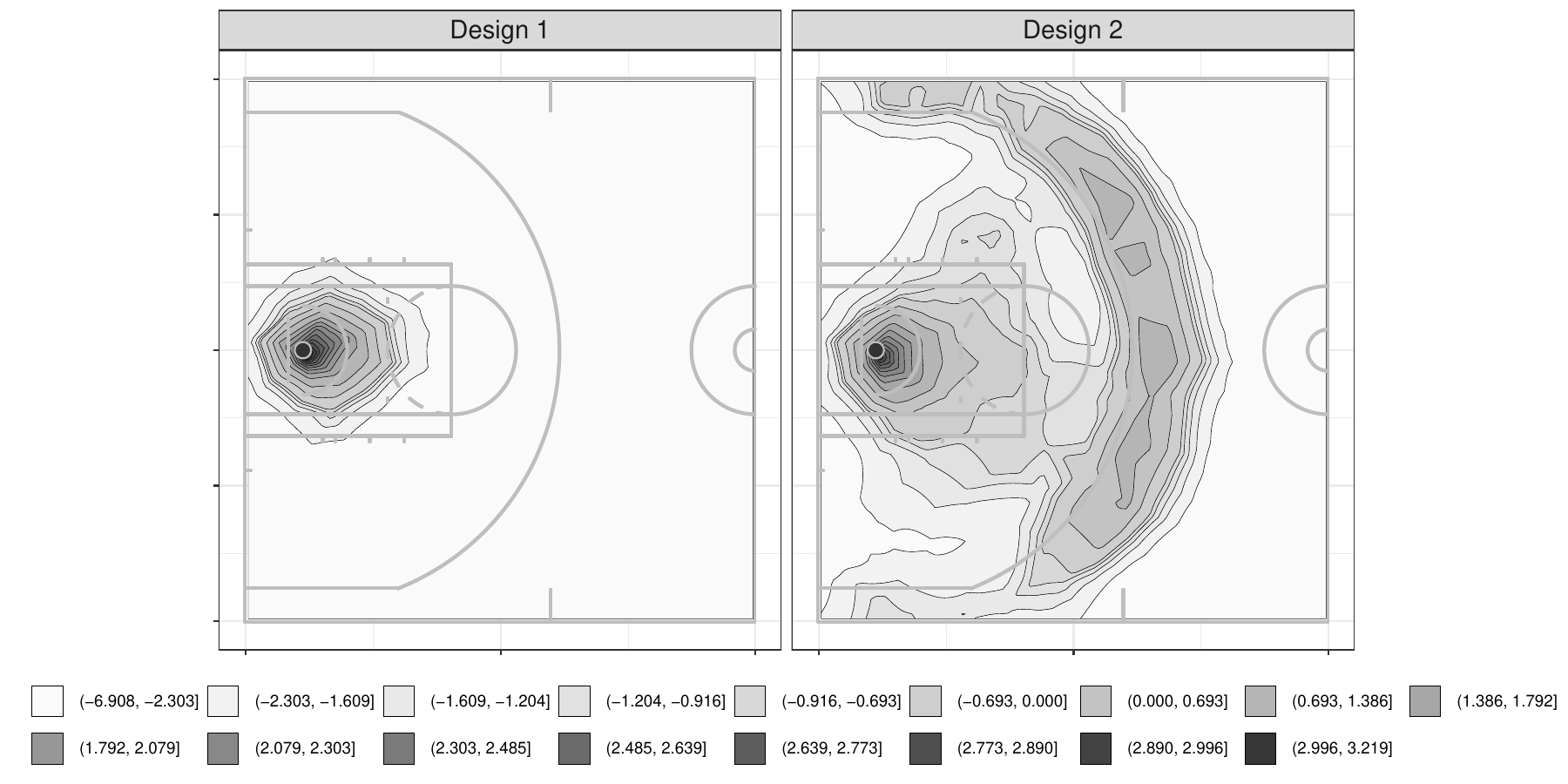}
    \caption{Contour Plots for Intensity Maps of Two Simulation Designs}
    \label{fig:depth_design}
\end{figure}

In this section, we randomly generate two groups of spatial point process realizations, denote as $\mathbb{G}_1=\{\mathbf{S}_1,\mathbf{S}_2,\cdots,\mathbf{S}_{100}\}$ and $\mathbb{G}_2=\{\mathbf{S}'_1,\mathbf{S}'_2,\cdots,\mathbf{S}'_{100}\}$.
Each group contains $100$ realizations; realizations in $\mathbb{G}_1$ follow a Poisson process with intensity map $\lambda_1(s)$, and realizations in $\mathbb{G}_2$ follow a Poisson process with intensity map $\lambda_2(s)$. The intensity maps are the discretized versions of intensity function, which can be denoted as a $112 \times 112$ gird. For each grid, the associated number stands for the total intensity of the grid. The log intensity maps of $\lambda_1$ (Design 1) and $\lambda_2$ (Design 2) are shown as Figure~\ref{fig:depth_design}.

Those two intensity maps which are calculated by \textsf{R}-package \textbf{lgcp} \citep{taylor2013lgcp} are picked from two representative NBA players (Steve Adams and James Harden). The first pattern corresponds to players most of whose shots are in the
painted area and another pattern represents shooting locations at the three-point line and inside the painted area. Under our simulation settings, the locations of field goal attempts are generated via the \textsf{R}-package \textbf{spatstat} \citep{baddeley2005spatstat} based on intensity maps shown in Figure~\ref{fig:depth_design}. As we can see from the above intensity maps, the intensity surfaces of Group 1 and Group 2 are different visually.

As the first step, we test to see if all test schemes could differentiate realizations in $\mathbb{G}_1$ from $\mathbb{G}_2$. We randomly select one realization from Group 1 and Group 2 separately (denoted as $\mathbf{S}$ and $\mathbf{S'}$) without replacement for $100$ times, and form $100$ hypothesis tests with the null hypothesis
$$H_0: \lambda(s;\mathbf{S})=\lambda(s;\mathbf{S'}),$$
or equivalently, $H_0: F(s;\mathbf{S})=F(s;\mathbf{S'})$, where $F(s;\mathbf{S})$ and $F(s;\mathbf{S'})$ are the corresponding distribution function of $\mathbf{S}$ and $\mathbf{S'}$. Here, we use the conventional significant level $5\%$ as criteria, and all $100$ tests are rejected for all test schemes (Method 1-Method 6). This result is expected since the two intensity functions are very distinct. Then we focus on testing performances regarding controlling type I error.

To estimate the type I error of different test schemes, we randomly select $2$ realizations from each group for $485$ times, then test against each other. The null hypothesis is $H_0: F(s;\mathbf{S})=F(s;\mathbf{S'})$, where $\mathbf{S}$ and $\mathbf{S'}$ sampled from the same group. We still use $5\%$ as the significant level and count the rejection number of different test schemes. The results are shown in Table \ref{table:R1}.

\begin{table}[htp]
\caption{Number of rejected tests for each group} 
\centering
\begin{tabular}{c c c c }
\hline\hline 
Method & Design 1 & Design 2  & Sum\\ [0.5ex] 
\hline % inserts single horizontal line
Liu-Singh's test on Cartesian coordinate (Method 1)&51&86&137\\
Liu-Singh's test on Polar coordinate (Method 2)&43&70&113\\ 
Two-dimensional KS test on Cartesian coordinate (Method 3)&52&55&110\\
Two-dimensional KS test on Polar coordinate (Method 4)&36& 52&88\\  
Our proposed test on Cartesian coordinate (Method 5)&\bf{23}&45&68\\
\bf{Our proposed test on Polar coordinate (Method 6)}&\bf{25}&\bf{25}&\bf{50}\\[1ex]
\hline 
\end{tabular}
\vspace{1ex}

{\raggedright   Note: There are 485 tests performed for each design based on methods we mentioned above. In total, for each method, we performed 970 tests.\par}
\label{table:R1}
\end{table}

From results shown in Table~\ref{table:R1}, we see that our proposed test (Method 6) is closest to the $5\%$ rejection boundary for all $970$ tests, which achieved the lowest type I error rate. We believe this is due to the combined effect of the two-dimensional KS test, polar system, and adoption of depth functions. If we compare the depth-based Liu-Singh's tests (Methods 1 and 2) with the two-dimensional KS tests (Methods 3 and 4), the two-dimensional KS tests perform better than Liu-Singh's tests, especially for Design 2. The partial reason is that Liu-Singh's test focuses more on data ranks and does not utilize the complete information of data.  
Similarly, comparing the tests on the Cartesian system (Methods 1, 3, and 5) with tests on the polar coordinate (Methods 2, 4, and 6) in Table~\ref{table:R1}, we notice that the usage of the polar system improves the overall test performance on controlling type I error.

In addition, we compare the two-dimensional KS tests (Methods 3 and 4) with our proposed tests (Methods 5 and 6). Although all these tests are based on the two-dimensional KS test, one can see that the type I errors are lower after adopting depth transformation (especially for Group 2). It further suggests that our proposed test outperformed other tests we considered in controlling type I error.

\subsection{Performance on power of test}

In this section, we emphasize on examining the sensitivity of our proposed test schemes with respect to small perturbations, i.e., the sensitivity of power with respect to the magnitude of shifts in distribution. This section aims to examine if the adoption of the polar system and depth function will improve the power of the test. Hence, we will mainly consider three different KS test schemes for comparison:

\begin{itemize}
\item The two-dimensional KS test (Method 3).

\item The two-dimensional KS test on the polar coordinate system (Method 4).

\item Our proposed depth-based test scheme (Method 6).
\end{itemize}

First, for each realization $\mathbf{S}_i=(s_{i,1},\cdots, s_{i,n_i})$ and  $\mathbf{S}’_i=(s’_{i,1},\cdots, s’_{i,m_i})$, $i=1,\cdots,100$ in $\mathbb{G}_1$ and $\mathbb{G}_2$ simulated previously, we add a random location-shift noise $\xi \sim Gaussian([0,0],0.25r I)$ on each location.  That is:
$$\mathbf{SS}_i=(s_{i,j}+\xi, j=1,\cdots,n_i),$$
$$\mathbf{SS}’_i=(s’_{i,j}+\xi, j=1,\cdots,m_i),$$
where $n_i$ and $m_i$ are the cardinalities of realization $\mathbf{S}_i$ and $\mathbf{S}’_i$, and $I$ is the two-dimensional identity matrix. Here we introduce a scale parameter $r$ to adjust the magnitude of the noise; as $r$ gets larger, the shifted distribution is more separated from the original distribution. 

\begin{figure}[htp]
\centering
  \begin{subfigure}{.48\textwidth}
  \centering
   \includegraphics[width=\textwidth]{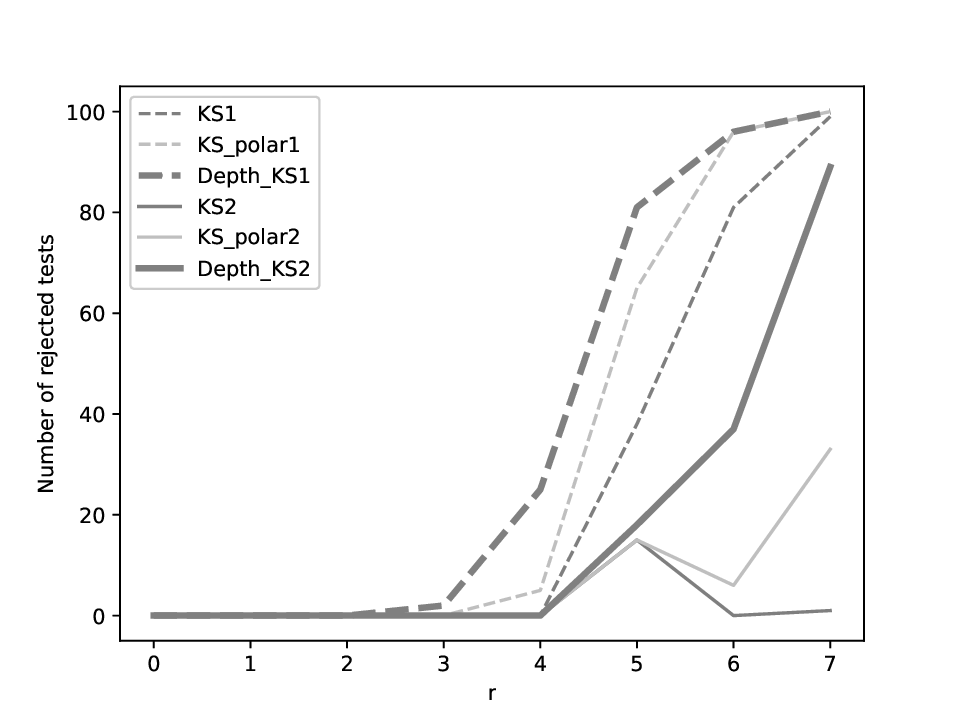}
   \caption{Number of rejected tests}
   \label{}
   \end{subfigure}
  \hfill
  \begin{subfigure}{.48\textwidth}
  \centering
   \includegraphics[width=\textwidth]{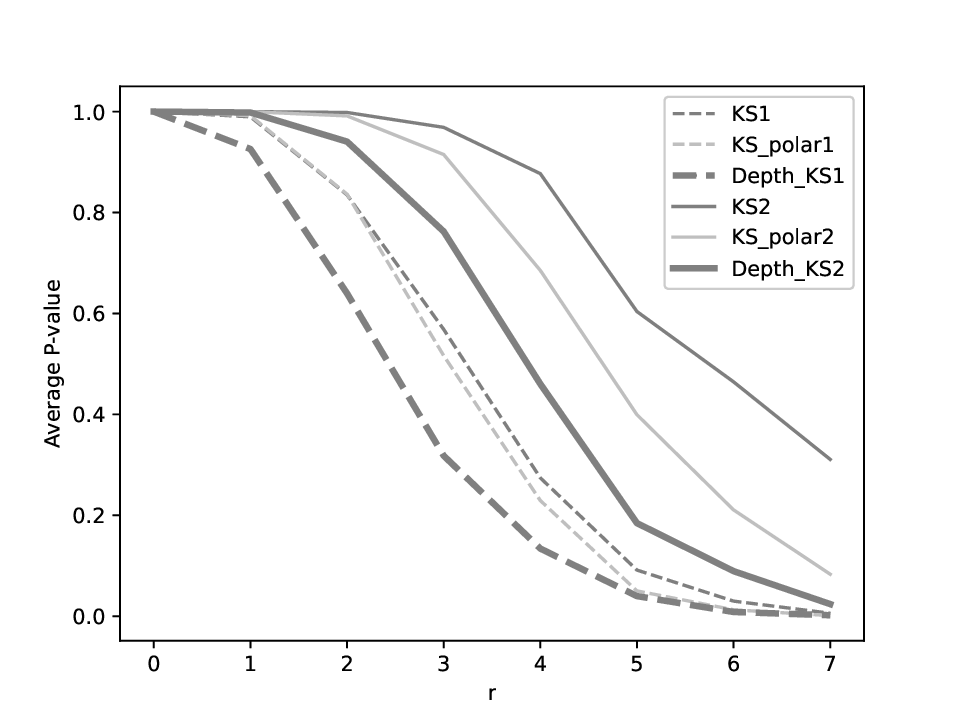}
   \caption{Average P-value}
   \label{}
   \end{subfigure}
   \caption{"KS1" is the two-dimensional KS test on Group 1 (Method 3), "KS-polar1" is the two-dimensional KS test based on polar coordinate for Group 1 (Method 4), and "Depth-KS1" is the proposed test scheme for Group 1.  Similar notation applies to Group 2. (a). Plot of number of rejected tests ($\alpha=0.05$) for Group 1 (dash line) and Group 2 (solid line) based on the proposed test scheme, KS test, and KS test on polar coordinate. (b).  Corresponding average P-value plot.}
   \label{fig:scale_shift}
\end{figure}

Note that the noise $\xi$ will only change the conditional distribution of location $F(\mathbf{S}_i||\mathbf{S}_i|=n_i)$, and the number of points (cardinality) $n_i$ remains untouched.
For each group, we can perform $100$ hypothesis tests with null hypothesis
$$H_0: F(\mathbf{S}_i)=F(\mathbf{SS}_i) \ \text{for} \ i \in [1,\cdots, 100],$$
or
$$H_0: F(\mathbf{S}'_i)=F(\mathbf{SS}'_i) \ \text{for} \ i \in [1,\cdots, 100].$$
By setting scale parameter $r$ to different values in $[0,1,2,3,4,5,6,7]$ and calculating the average p-value of $100$ tests, we have the result in Figure \ref{fig:scale_shift}.
From Figure \ref{fig:scale_shift}, we see that our proposed test is more sensitive to small Gaussian noises than the traditional KS test or KS test on the polar system. The average P-values of our proposed KS test scheme decrease more rapidly than the other two tests, suggesting that the adoption of polar coordinate and depth function could improve the test power.

In addition, we add tiny variations directly on the intensity map, regenerate $100$ realizations, and then test against the original realizations. As we mentioned early, the intensity map can be viewed as a $112 \times 112$ grid, and the number in each grid stands for the total intensity of the corresponding area. For each grid, we add a noise term $\epsilon \sim F(r)$, where $F(r)$ is a distribution with a parameter $r$ to adjust the magnitude of noise.

The distributions of the noise term we consider are
\begin{itemize}
    \item Half-Gaussian distribution:  $\epsilon \sim |N(0,r)|$, 

    \item Lognormal distribution $\log(\epsilon) \sim N(0,r)$.
\end{itemize}
By adjusting $r$ from $[0.025,0.05,0.075,0.125,0.15]$ for the Half-Gaussian noise and $[0.1, 0.5, 0.75, 1, 1.25]$ for the Lognormal noise, we get the test results shown in Table \ref{table:R2}.
\begin{table}[htp]
\caption{Number of rejected tests out of 100 tests}
\centering
\begin{tabular}{ |c|c|c|c|c| }
%\hline
%\multicolumn{5}{ |c| }{Number of rejected tests out of 100 tests} \\
\hline
Type of noise & r & KS (Method 3) & KS polar(method 4) & The proposed test (Method 6)\\ \hline
\multirow{5}{*}{$\epsilon\sim |N|$} & 0.025 & $59 \mid 9$ & $73 \mid 12$ & $57 \mid 15$ \\
													  & 0.05 & $100 \mid 13$ & $100 \mid 27$ & $100 \mid 29$ \\
													  & 0.075 & $100 \mid 23$ & $100 \mid 37$ & $100 \mid 42$\\
 													  & 0.125 & $100 \mid 72$ & $100 \mid 94$ & $100 \mid 91$\\
 												      & 0.15 & $100 \mid 93$ & $100 \mid 100$ & $100 \mid 99$\\ \hline
\multirow{5}{*}{$\log(\epsilon) \sim N$} & 0.1 & $6 \mid 5$ & $2 \mid 3$ & $12 \mid 5$  \\
                                                           & 0.5 & $8 \mid 8$ & $8 \mid 4$ & $13 \mid 8$ \\
                                                           & 0.75 & $23 \mid 11$ & $23 \mid 15$ & $25 \mid 20$ \\
                                                           & 1 & $46 \mid 34$ & $42 \mid 34$ & $57 \mid 41$ \\
                                                           & 1.25 & $96 \mid 94$ & $99 \mid 99$ & $99 \mid 100$ \\ 
\hline
\end{tabular}
\vspace{1ex}

{\raggedright  Note: The test results are based on criteria $\alpha=0.05$. The number in the front of separate '$|$' is the number of rejected tests for Group 1, the number after '$|$' is the number of rejection for Group 2, and each group contain 100 tests.  \par}
\label{table:R2}
\end{table}

Finally, we shift the intensity map by $c$ pixel horizontally, regenerate $100$ realizations according to the shifted intensity map, and test against the original realizations. Since the intensity map of Group 1 ($\lambda_1$) is too concentrated and sensitive to small location shifts, all tests are rejected after 1 pixel shift. Here we discard tests results for $\lambda_1$ and only test $\lambda_2$ vs. $\lambda^r_2$ (the shifted intensity map by $r$ pixel). The results are shown in Table \ref{table:R3}.

\begin{table}[H]
\centering
\caption{Number of rejected tests out of 100 location-shift tests}
\begin{tabular}{ |c|c|c|c| }
%\caption{Number of rejected tests out of 100 tests}
\hline
Shift by $c$ pixel & KS (Method 3) & KS polar (Method 4) & The proposed test (Method 6)\\ \hline
1 & 24 & 30 & 36 \\
2 & 76 & 82 & 85\\
3 & 100 &100 & 100\\
4 & 100 & 100 & 100\\
5 & 100 & 100 & 100\\ \hline
\end{tabular}
\vspace{1ex}

{\raggedright  Note: The test results are based on criteria $\alpha=0.05$. \par}
\label{table:R3}
\end{table}

From Table \ref{table:R2}, one can see that for Half-Gaussian noises, our proposed test scheme is more sensitive to small noises in intensity maps, especially in Group 2 ($\lambda_2$). This indicates our proposed test has the ability to identifying small differences among underlying intensity maps. Although the test results for Method 4 and Method 6 are similar, the slight improvement suggests the adoption of depth transformation is suitable when dealing with Half-Gaussian noises. The test results for Lognormal noises further support our supposition, for which the improvement after adopting depth transformation is more evident by comparing Method 6 with Method 3 and 4.

In contrast in Table \ref{table:R3}, the test results suggest both polar and depth transformations will improve the sensitivity of the test. By comparing Method 3 and Method 4, the only difference is whether the KS test on polar coordinates are used, and the enhancement is solely based on the implementation of the polar system. On the other hand, the improvement in sensitivity from Method 4 to Method 6 attributes to the usage of depth transformation, which is the only difference between these two methods.

\section{Professional Basketball Data Analysis}\label{sec:app}
\subsection{Data Overview}
Our data consists of both made and missed field goal attempt locations from the offensive half court of games in the 2017-2018 National Basketball Association (NBA) regular season. The visualization of selected players is shown in Section~\ref{sec:data}. In this section, we focus on players who have made more than 400 field goal attempts. Also, the rookie players in that season, such as Lonzo Ball and Jayson Tatum, are not considered. In total, we have 191 players who meet the two criteria above in our analysis. 
\subsection{Hypothesis Testing for NBA Players}
\begin{figure}[htp]
    \centering
    \includegraphics[width=\textwidth]{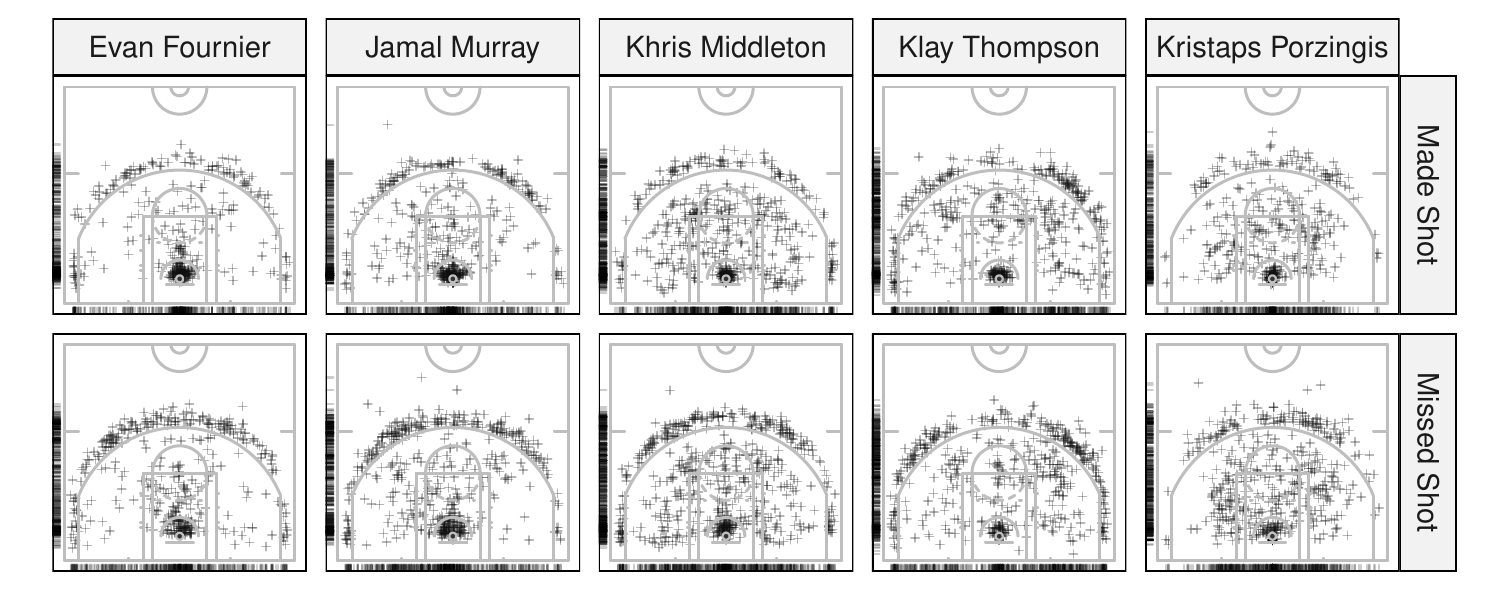}
    \caption{Made (Top) and missed (Bottom) shot locations of five example players (Evan Fournier, Jamal Murray, Khris Middleton, Klay Thompson, Kristaps Porzingis) whose missed and made shot patterns are not significantly different (with $\alpha=0.05$) by our proposed test, but are different by the other two methods (KS test and KS test after Polar transition) in our comparison.}
    \label{fig:real1}
\end{figure}

\begin{figure}[htp]
    \centering
    \includegraphics[width=\textwidth]{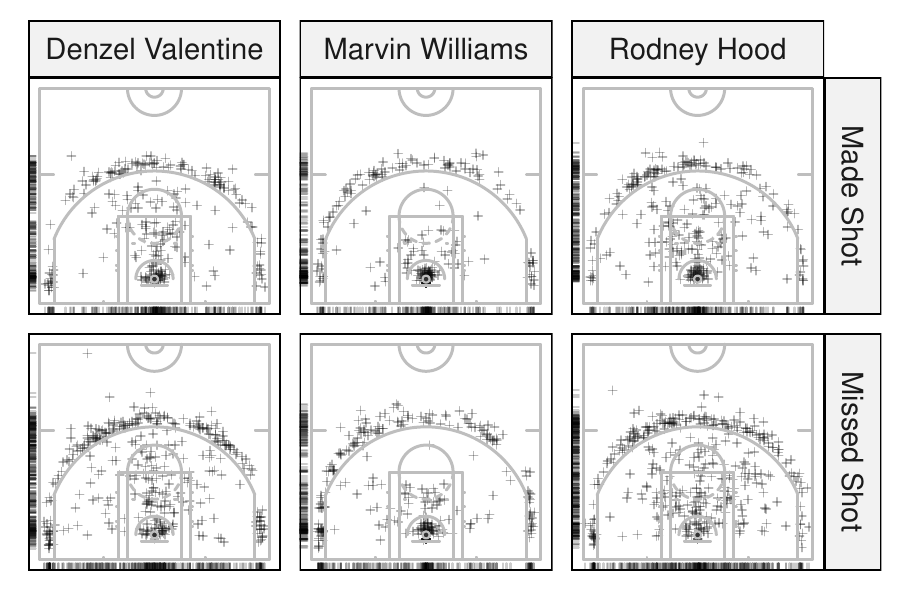}
    \caption{Made (Top) and missed (Bottom) shot locations of three example players (Denzel Valentine, Marvin Williams, Rodney Hood) whose missed and made shot patterns are significantly different (with $\alpha=0.05$) by our proposed test, but are not different by the other two methods (KS test and KS test after Polar transition) in our comparison.}
    \label{fig:real2}
\end{figure}

Our goal is to discriminate the difference between made process and missed process for NBA players. This section also includes the KS test and KS test with polar coordinates in contrast to our proposed depth-based test for all 191 players. Our testing procedure rejects the null hypothesis for 147 players under $\alpha=0.05$ and fails to reject the null hypothesis for 44 players. However, the other two methods fail to reject the null hypothesis for just 13 players. The details of players' names for the three testing procedures are shown in supplementary materials. In order to have a closer look at our testing results, we pick several representative players to visualize their made and missed locations in Figure~\ref{fig:real1} and Figure~\ref{fig:real2}.

Figure~\ref{fig:real1} contains 5 players. Our testing procedure does not reject the null hypothesis but the other two testing procedures reject the null hypothesis for these 5 players. From this figure, we see that the missed and made shot patterns for these players are very similar.

Figure~\ref{fig:real2} contains 3 players. Our testing procedure rejects the null hypothesis but the other two testing procedures do not for these 3 players. Clearly, the missed and made shot patterns of these 3 players are quite different: Hood has more missed shots on top 3 pointers; Valentine has different patterns on long 2-point; Williams has more missed on the right and left top 3 pointers.

\begin{figure}
    \centering
    \includegraphics[width=\textwidth]{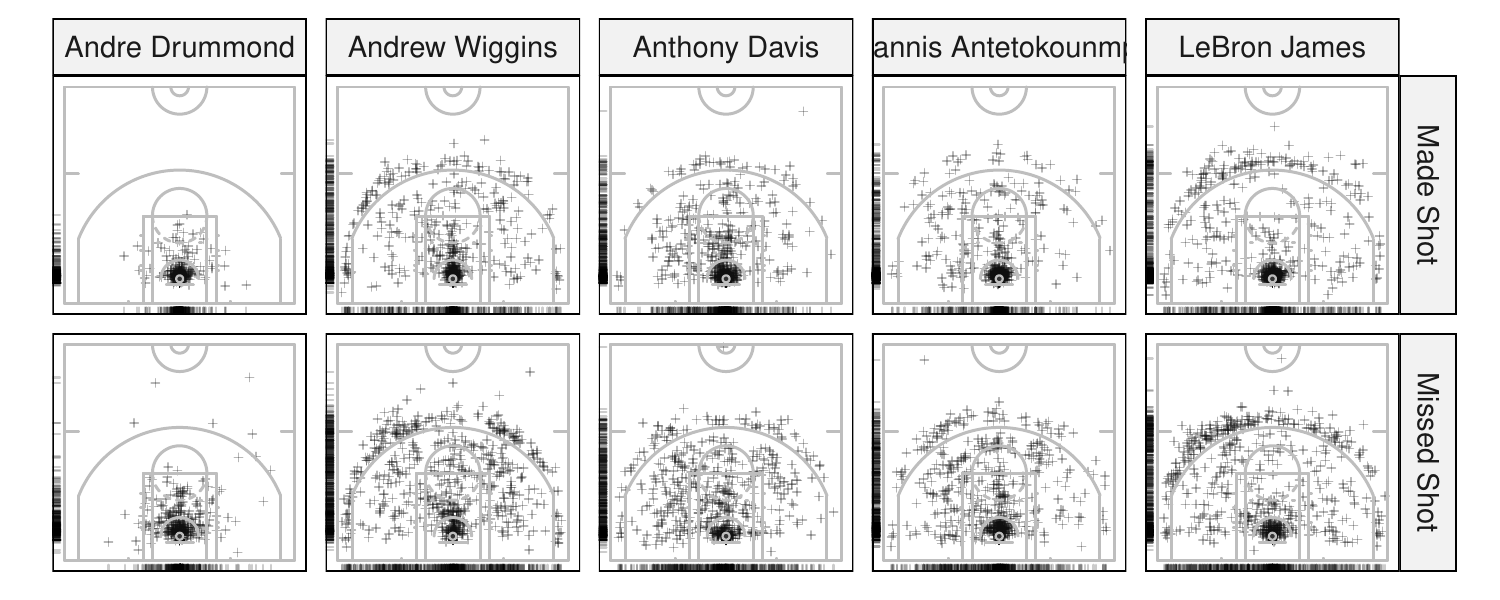}
    \caption{Top 5 Players with smallest P-values based on our proposed test framework. Each column indicates the Made pattern (Top) and Missed pattern (Bottom) of the respective player. }
    \label{fig:real3}
\end{figure}

In addition, we have the top 5 players (Giannis Antetokounmpo, LeBron James, Anthony Davis, Andrew Wiggins, Andre Drummond) with the smallest p-value in Figure~\ref{fig:real3}. We see that the listed five players' made processes are much different from their missed processes. For example, Andre Drummond does not have successful FTAs out of paint region, and Antetokounmpo has many more missed shots on right-wing 3-pt and long 2-pt.

Furthermore, we pull the top 5 players with the highest field goal attempts of each position (C, PF, SF, SG, and PG). The three testing procedures have consistent results, rejecting the null hypothesis at $\alpha=0.05$ for all five positions. The results are shown in supplementary materials.

In a brief summary, for most players (nearly 77\% players in our analysis), the made process and the missed process are different based on our testing procedures. This means most players have their hot zones and cold zones on the court. The best defense strategy is to force those players to shoot more on their cold zones than hot zones. For the offense side, the coach needs to design offensive tactics such as screening to make their players have more comfortable shots during the game.

\subsection{Players Classification}
Naturally, a significant test and P-value can be applied to classify point processes as it measures how extremely different one point process is from another under the null hypothesis that the two processes are the same. In this section, we will apply our proposed test framework to classify different players into different groups. The test steps can be generalized as follows:

\begin{enumerate}
\item Randomly choose one player, and use his missed and made pattern as the benchmark;
\item Calculate the paired-wise P-value of made and missed pattern of the benchmark player with all other players, denote as $P_{missed}(player1, player2)$ and $P_{made}(player1, player2)$;
\item Compare the calculated P-values with a pre-defined significant level(threshold), e.g., 0.025. If both $P_{missed}(player1, player2)$ and $P_{made}(player1, player2)$ are above a certain threshold, then classify it to the benchmark player's group. Since we are doing multiple testing, i.e., comparing made and missed simultaneously, we applied the Bonferroni Correction to adjust the significance level $\alpha_{adj}=0.05/2=0.025$. 
\item Randomly choose another player as threshold, and repeat step 1)-3).
\end{enumerate}

Based on the procedures above, the classification results of 5 selected players are given in Table \ref{tab:classfication}.
\begin{table}[htp]
    \centering    \caption{Classification results of 5 selected players}
\begin{tabular}{|l|p{110mm}|}
\hline
Benchmark Player & Players in Same Group\\
\hline
DeAndre Jordan & N/A\\
\hline
DeMar DeRozan & Rondae HollisJefferson, Cory Joseph\\
\hline
Giannis Antetokounmpo & N/A\\
\hline
LeBron James & Eric Bledsoe, Aaron Gordon, Jeff Green, James Johnson, Mario Hezonja, DAngelo Russell, Andrew Harrison \\
\hline
Stephen Curry & Jaylen Brown, Terry Rozier, Wilson Chandler, Trey Lyles, Reggie Bullock, DJ Augustin, Stanley Johnson, Manu Ginobili, Justise Winslow, Lou Williams\\
\hline
\end{tabular}

    \label{tab:classfication}
\end{table}
Based on results shown in Table~\ref{tab:classfication}, we see that Giannis Antetokounmpo and DeAndre Jordan are two special players we can not find any players with similar missed and made patterns. In addition, LeBron James's shooting pattern is more like Power Forward.

\subsection{LeBron James V.S. Stephen Curry}
Now, we will address the initial question we posed: Do made and missed field goal attempts follow different spatial processes for Stephen Curry? Additionally, we will explore whether LeBron James has a higher number of field goal attempts in the region where he has a higher field goal percentage.

The answer to the first question is straightforward. Based on the KS test, KS test after polar transformation, and our proposed test framework, all indicate that the patterns of made and missed shots for Stephen Curry differ significantly at the 5\% significance level. This suggests that he has specific shooting areas where he achieves a higher success rate compared to misses. However, at the 1\% significance level, our proposed test (as well as the KS test) would reject the null hypothesis, implying that his preferred shooting spots may not be as strong or as numerous. Considering the injury he sustained this season, this outcome is somewhat expected.

Regarding the second question, all the tests in our comparison indicate that LeBron James' made patterns and missed patterns are statistically different. This suggests that in certain areas, he has a higher success rate in made shots compared to misses, or vice versa. When we examine the made and missed patterns in Figure \ref{fig:real3}, it becomes evident that he attempts more field goals in the upper left three-point line and has a higher success rate in that area as well. However, around the upper right three-point line and the free-throw area, his success rate is less satisfactory, resulting in numerous missed attempts.

Furthermore, based on the classification results in Table \ref{tab:classfication}, we can offer some suggestions to coaches who aim to train specific types of players or compete against them. For example, we observed that LeBron James has similar shooting and missed shot patterns to Eric Bledsoe, Aaron Gordon, Jeff Green, James Johnson, Mario Hezonja, D'Angelo Russell, and Andrew Harrison. From a coaching perspective, this suggests that those players could benefit from a similar training plan as LeBron James, who was recognized as one of the most successful players in the 2017-2018 NBA regular season and has had the highest Points Per Game (PPG) in the past 10 years. Additionally, if there is an effective defensive plan against players like LeBron James, it could apply to other players as well.
\section{Discussion}\label{sec:disc}

In this paper, we propose a depth-based testing procedure for discriminating two spatial point processes. Building upon the polar coordinate system, we develop a two-dimensional KS test on one-dimensional depth. Unlike several benchmark test procedures, our proposed methods apply a multivariate KS test with a one-dimensional depth function. The depth function then acts like a Box-Cox transformation, transforming data into a different shape rather than reducing dimensions. By comparing with several benchmark testing procedures, the numerical results showed that the proposed method can provide a robust testing procedure under both null and alternative hypotheses. In the analysis of the NBA shot charts data, our proposed method reveals several interesting findings for different NBA players and hence provides more \emph{objective} and principled analysis for shooting patterns of the players in NBA. Compared with traditional point process methods for shot charts of NBA players, our proposed testing procedures successfully discriminate the differences between made and missed shots of NBA players. For some top players such as Anthony Davis, Lebron James, and Giannis Antetokounmpo, they have smallest p-values. The comfortable zones for those players are very obvious. In addition, we can see that Giannis Antetokounmpo is a very distinct player in NBA. We cannot find any players who have similar missed and made pattern with him. 

Unlike previous studies that have focused on goal attempt frequency and accuracy of specific players (e.g., \citet{brian06}), our test framework aims to determine whether two shot patterns differ from each other while also capturing the spatial randomness of field goal attempts. Instead of using complex models to analyze the relationships between defenders and offenders or estimating shoot intensity maps (e.g., \citet{miller2014factorized}), we have developed a simple yet powerful statistical test framework to differentiate spatial patterns based on the center-outward tendency of players' field-goal attempts. 

Our framework significantly differs from the methodologies discussed in the introduction section. It can effectively summarize the overall differences in shot patterns, making it particularly valuable from coaching and recruiting perspectives. By using this approach, teams can identify and hire players with desired shot patterns, and its potential in classification allows recruiters to quickly identify groups of players with similar shot patterns.

However, it also comes with certain limitations. First, our proposed test framework primarily aims to determine if one spatial pattern significantly differs from another. But, without additional information, such as performance-related summary statistics of players, our model lacks the ability to identify which shooting pattern is superior. Furthermore, in this research, we made the assumption that the shot locations adhere to a Poisson distribution. This assumption stems from the belief that shot locations are independent of previous locations, which unfortunately prevents us from capturing the temporal relationships between shot positions of teammates or defenders, as explored in other studies on spatio-temporal processes (e.g., \cite{Daniel2016} proposed a framework based on Gaussian processes to model the evolution of a basketball possession). Additionally, our analysis using shot chart data is incapable of accounting for complicated time dependence, in contrast to studies such as \cite{Franks2015} and \cite{Daniel2016} that utilize real-time tracking data. Therefore, our model is unable to comprehend the influence of defenders on the shot effectiveness and correlations between successful and missed shot attempts. 

Some topics beyond the scope of this paper are worth further investigation. First, an extension of the proposed testing framework to more than two-dimensional spatial point processes will provide broader applicability. Second, developing a depth-based testing procedure for pair correlation functions of point processes is another exciting topic. Moreover, false discovery rate control in multiple testing problems represents an interesting direction for future work. From the application point of view, having a closer look at court locations with the most significant effects will provide more data-driven information for NBA teams.

\bibliographystyle{chicago}
\bibliography{main}
\end{document}